\documentclass{sig-alternate}
\usepackage{times}
\usepackage{graphicx}
\usepackage{cite}	
\usepackage{url}

\long\def\com#1{}

\long\def\xxx#1{{\bf XXX: }{\small [#1]}}

\title{Determinating Timing Channels in Compute Clouds}
\numberofauthors{2}
\author{
	\alignauthor
	Amittai Aviram, Sen Hu, Bryan Ford \\
	\affaddr{Yale University}
	\alignauthor
	Ramakrishna Gummadi \\
	\affaddr{University of Massachusetts Amherst}
}

\begin{document}

\maketitle

\begin{abstract}
Timing side-channels
represent an insidious security challenge for cloud computing,
because:
(a) massive parallelism in the cloud
makes timing channels pervasive and hard to control;
(b) timing channels enable one customer to steal information from another
without leaving a trail or raising alarms;
(c) only the cloud provider can feasibly detect and report such attacks,
but the provider's incentives are {\em not} to; and
(d) resource partitioning schemes for timing channel control
undermine statistical sharing efficiency,
and, with it, the cloud computing business model.
We propose a new approach to timing channel control,
using {\em provider-enforced deterministic execution}
instead of resource partitioning
to eliminate timing channels within a shared cloud domain.
Provider-enforced determinism prevents execution timing
from affecting the results of a compute task, however large or parallel,
ensuring that a task's outputs leak no timing information
apart from explicit timing inputs and total compute duration.
Experiments with a prototype OS for deterministic cloud computing
suggest that such an approach may be practical and efficient.
The OS supports deterministic versions of familiar APIs
such as processes, threads, shared memory, and file systems,
and runs coarse-grained parallel tasks
as efficiently and scalably as
current timing channel-ridden systems.
\end{abstract}

\section{Introduction}

It is hotly debated whether individuals and companies
should trust cloud providers with 
sensitive information, 
but few would suggest that a cloud customer should
trust the provider {\em and} all the provider's other customers.
Yet this may soon be the cloud's {\em de facto} security model---%
if it isn't already---%
due to timing channels.

\com{
Cloud services are riddled with classic covert timing channels~\cite{
	kemmerer83shared,wray91analysis},
allowing subversion of information flow control~\cite{
	efstathopoulos05labels,zeldovich06making}.
Further, many timing channels
permit one computation to steal secrets 
from {\em non-cooperating} processes sharing the same hardware~\cite{
	aciicmez07predicting,aciicmez07yet,percival05cache,wang06covert},
including specifically in the cloud~\cite{ristenpart09cloud}.
}

Timing channels are well-known and well-studied~\cite{
	kemmerer83shared,wray91analysis},
originally driven by military-grade security demands.
They have gained broader relevance, however,
in the context of commercially applicable
information flow control~\cite{
	efstathopoulos05labels,zeldovich06making},
and due to the discovery that computations
{\em unintentionally} broadcast sensitive information
via numerous timing channels in shared environments.
A sensitive computation sharing a CPU core with an attacker,
through either time division or hyperthreading,
is akin to standing behind a transparent shower door:
e.g., an attacker may steal information from the victim via
the shared L1 data cache~\cite{percival05cache},
shared functional units~\cite{wang06covert},
the branch target cache~\cite{aciicmez07predicting}, or
the instruction cache~\cite{aciicmez07yet}.

Most of the above attacks were demonstrated
between processes on a conventional OS,
but per-customer VMs on a provider-owned machine
share resources in essentially the same way,
making the results theoretically applicable to clouds---%
especially those relying on ``container-based'' virtualization~\cite{
	soltesz07container}.
Timing attacks have even been demonstrated specifically on VMs
commonly used in clouds~\cite{ristenpart09cloud},
although it is not yet clear how easily these lab-based experiments
could be replicated in a noisy commercial cloud.

Whether timing channels represent
an immediate security threat
or merely a hairline fracture,
it is worth repeating the security adage,
``attacks never get worse; they only get better.''
Today's timing-channel exploits pick low-hanging fruit,
extracting information from only one high-bandwidth timing channel at a time
via straightforward analysis techniques.
Shared computing environments have
many other timing channels,
such as L3 caches shared between cores,
memory and I/O busses,
and cluster interconnects.
There are probably ways to extract weaker signals from stronger noise,
aggregate information from low-rate leaks over time,
correlate leaks across multiple channels, etc.
Attack amplification techniques applicable to arbitrary timing channels
have already appeared~\cite{potlapally06satisfiability}.
It would simply be foolish for us to expect timing attacks
{\em not} to continue getting more effective and more practical over time.

In the rest of this paper,
we set aside the ``imminence of threat'' debate
and simply assume that at {\em some} point, sooner or later,
timing channels will become an important cloud security issue.
We focus here on understanding
the basic nature of the timing channel problem in the cloud context,
independent of specific channels and attacks,
and on discovering potential solutions compatible with
the requirements of cloud environments.
\com{
We first identify ways
in which cloud computing amplifies timing channel risks,
then propose a general technique to control timing channels
within a compute/storage cloud,
and finally present preliminary evidence of the technique's viability
for certain broad classes of cloud applications.
}
We focus in particular on timing channels {\em internal} to a cloud:
other side-channels,
such as those derived from a client's communication
with a cloud-based service~\cite{chen10side},
are also important but beyond our present scope.

We make three main contributions.
First, we identify four ways
the cloud computing model amplifies timing channel security risks
compared with traditional infrastructure.
Second, we propose a new method of timing channel control
based on provider-enforced deterministic execution,
which aggregates {\em all} internal timing channels
into a single controllable channel at the cloud's border.
Third, we present a proof-of-concept cloud computing OS that enforces
determinism, with preliminary results 
suggesting that it could support parallel cloud applications efficiently
without sacrificing
the cloud provider's flexibility in allocating resources to clients.

\com{
We make three main contributions.
First, we identify three specific ways
in which cloud computing amplifies timing channel security risks
compared with traditional private infrastructure.
Second, we propose a novel (to our knowledge) cloud computing architecture
based on provider-enforced deterministic execution,
which closes {\em all} internal timing channels, regardless of type,
within a shared execution domain,
leaving only one easily-controlled timing information leak
at the domain border.
Third, we present a proof-of-concept cloud computing OS
and preliminary results
suggesting that provider-enforced determinism
could offer both convenience to developers of parallel cloud applications
and sufficient efficiency and scalability for
at least some parallel applications.
}

\com{
Compared with traditional private infrastructure,
cloud computing amplifies timing channel security risks
in at least three ways.
First and most obviously,
because unrelated and mutually untrusting customers
may concurrently share computing hardware,
timing attacks that would be practical only for ``insiders''
on private infrastructure
become practical ``outsider'' attacks in the cloud.
Second, while private infrastructure owners
have both the prerogative and ability to monitor all software
running on their machines for malicious code,
cloud customers cannot monitor other customers' computations for attacks,
and cloud providers have neither incentive nor prerogative
to monitor their customers' computations for such attacks.
Third, the only known general methods of closing timing channels of all types
are to partition shared resources statically,
or inject noise by adding substantial random loads,
both of which undermine the cloud business model
by compromising the provider's ability to oversubscribe
and statistically multiplex shared hardware resources.
}

\com{
XXX do I want to keep writing the intro in this form,
or is it getting too long?  Maybe the short version is better?
XXX (AFA)  Yes, it was too long.
}

\com{ 
Section~\ref{sec-motiv}
explores the timing channel problem in the context of cloud computing,
Section~\ref{sec-arch}
presents our timing-hardened cloud architecture,
Section~\ref{sec-impl}
outlines our prototype cloud computing OS
and 
presents preliminary results,
Section~\ref{sec-related}
summarizes related work, and
Section~\ref{sec-concl}
concludes.
}

\section{Timing Channels in the Cloud}
\label{sec-motiv}

\com{ 
XXX (AFA) I don't think the point below is true.
See, for instance,
http://www.hh.se/download/18.3d00a1db126d3c114b9800075/8+Is+Cloud+Computing+Ready.pdf
XXX Clarify point: much of the current cloud privacy hoopla
is about whether we can trust the provider,
but timing channels means you're vulnerable not just to the provider
but to other customers sharing your cloud!} 

Current cloud privacy discussions focus on the provider's 
obligation to enforce security and
earn the customer's trust.
These discussions presuppose the provider's full awareness
of the security risks from which it must shield the 
customer~\cite{liu02analyzingsecurity,pearson09privacy}.
But exposure to malice from another customer's software may be hard
for the provider to detect or prevent without careful consideration
of the cloud's architecture.
Timing channels typify such insidious risks.



Although timing channels represent an important security risk
in any shared infrastructure,
the cloud model exacerbates these risks
in at least four specific ways, which we discuss below.
The first two points are well-known to some but worth repeating,
while to our knowledge the second two have not previously been discussed.

\com{
First, the business of cloud providers is to offer
a highly efficient, general-purpose programming environments
on which they are happy to run {\em any} computation
provided by a paying customer,
with ``no questions asked'' about what is being computed.
No cloud provider wants to tell a potential customer
that a particular application cannot be ``moved to the cloud''
because a particular hardware or operating system infrastructure feature
required by the application is unavailable there,
or because performance is too poor or unpredictable
when running under the cloud provider's virtual machine monitor (VMM).
}

\paragraph{Parallelism creates pervasive timing channels}

In the days of uniprocessors and single-threaded processes,
it was possible to control timing channels
by limiting untrusted processes' access to high-resolution clocks and timers,
and to other I/O devices that can behave like clocks~\cite{
	kemmerer83shared,wray91analysis}.
But today's increasingly parallelism-oriented hardware---%
especially in the massively parallel cloud context---%
creates numerous implicit, high-resolution clocks
that have nothing to do with I/O.
Hardware caches and interconnects in their many forms
all represent shared resources
that can be modulated~\cite{
	percival05cache,wang06covert,aciicmez07predicting,aciicmez07yet}.
A thread running in a loop
can create a high-resolution reference clock~\cite{wray91analysis},
as illustrated by the trivial code in Figure~\ref{fig-refclock},
even if the OS or VM has virtualized or disabled
all ``explicit'' hardware clocks.
Even processes
with no access to explicit clocks, timers, or other devices,
can thus use parallelism-derived implicit clocks to exploit timing channels.

\begin{figure}[t]
\centering
\begin{small}
\begin{verbatim}
volatile long long timer = 0;

void *timer_func(void *)
  { while (1) timer++; }

main() {
  pthread_create(&timer_thread, NULL,
                 timer_func, NULL);
  ...
  // Read the "current time"
  long long timestamp = timer;
  ...
}
\end{verbatim}
\end{small}
\caption{Implementing a high-resolution reference clock using threads,
	when no explicit hardware clocks are available.}
\label{fig-refclock}
\end{figure}

\paragraph{Insider attacks become outsider attacks}

With notable exceptions~\cite{brumley03remote},
timing channel exploits usually require the attacker to run
a sophisticated, CPU-intensive program on the victim's machine.
On private infrastructure, this usually means
the attacker must be an ``insider'' or have already compromised the machine.
But a cloud provider's business is to run any paying customer's computation
with ``no questions asked.''
Since the provider may colocate arbitrary customers' computations
on a given machine without the knowledge or consent of either,
a timing attack exploitable only by ``insiders'' on private infrastructure
may be mounted by malicious ``outsiders'' in the cloud.
An attacker may simply ``fish'' for secrets
without even knowing the identity of the co-resident victim,
by monitoring timing channels for SSH keystrokes for example,
or the attacker may deliberately attempt to obtain co-residency
with a specific target~\cite{ristenpart09cloud}.

\com{
The first property means that,
to the extent
a cloud provider succeeds in providing features and performance
equivalent to what a customer would get on private infrastructure,
a timing attack that one process or virtual machine
might be able to mount against another in a private infrastructure
is also likely to be portable to the cloud.
The second property means that many of these attacks,
which might be exploitable only by ``insiders'' on private infrastructure,
may be mounted by arbitrary ``outsiders'' in the cloud context.
To exploit timing channels via shared CPU resources, for example,
the attacker must be able to upload and run
a fairly sophisticated, CPU-intensive program
on the same machine as the victim's program.
The attacker is unlikely to be able to do this on private infrastucture
unless he has already seriously compromised the victim's machine.
If the victim program resides in the cloud, however,
the attacker can obtain his attack platform
merely by opening an account with the same provider
and paying for the required CPU time.
(The attacker may have to rent some time on many cloud nodes
in order to find the one(s) on which the victim's software is running,
but doing this search may not be that difficult or expensive
if the attacker knows some easily discernible ``timing signature''
produced the victim's software through {\em any} shared resource.)
}

\paragraph{Cloud-based timing attacks are unlikely to be caught}
The owner of private infrastructure has the right
to monitor and inspect any running software to detect malicious code.
Cloud customers cannot monitor other customers' computations
to protect themselves against timing attacks, however
(except by engaging in ``counter-espionage'' attacks themselves),
and cloud providers have no prerogative
to monitor their customers' computations
due to customer privacy concerns.
Since a timing attack leaves no trail of compromised protection mechanisms,
successful timing attacks are unlikely to raise alarms
and will probably just go unnoticed.
Thus, providers risk nothing
by leaving timing attacks undetected and unreported,
whereas monitoring customers in order to detect and report such attacks
may invite privacy lawsuits.

\paragraph{Controlling timing channels via resource partitioning
		undermines the cloud's elasticity and business model}
One general approach to controlling timing channels
is to limit the rate at which one user's demand for a shared resource
may visibly affect the resource's availability to another user,
either by statically partitioning the resource
or injecting noise into scheduling decisions.
Recent cache partitioning proposals
exemplify this approach~\cite{kong08deconstructing}.
These methods limit the provider's ability
to oversubscribe and statistically multiplex shared hardware efficiently
among users,
however,
undermining the basic business model of cloud computing.
Without statistical multiplexing,
the cloud loses its elasticity,
leaving the provider essentially selling
only private infrastructure hosting and outsourced management services.

\section{A Timing-Hardened Cloud}
\label{sec-arch}

We now explore a cloud computing architecture
that closes all internal timing channels,
regardless of number and types of shared resources,
leaving only one controllable timing channel at the boundary.
The basic idea is to make the cloud behave
like a deterministic batch job processor,
reminiscent of early mainframes.

A computation needs access to two ``clocks'' to exploit any timing channel:
a {\em reference clock}
and a clock that can be {\em modulated}~\cite{wray91analysis}.
While standard approaches to timing channel control
attempt to limit visible clock modulation,
our approach is to eliminate all internal reference clocks---%
even in the presence of parallelism.


\subsection{Provider-Enforced Determinism}

\begin{figure}[t]
\centering
\includegraphics[width=0.47\textwidth]{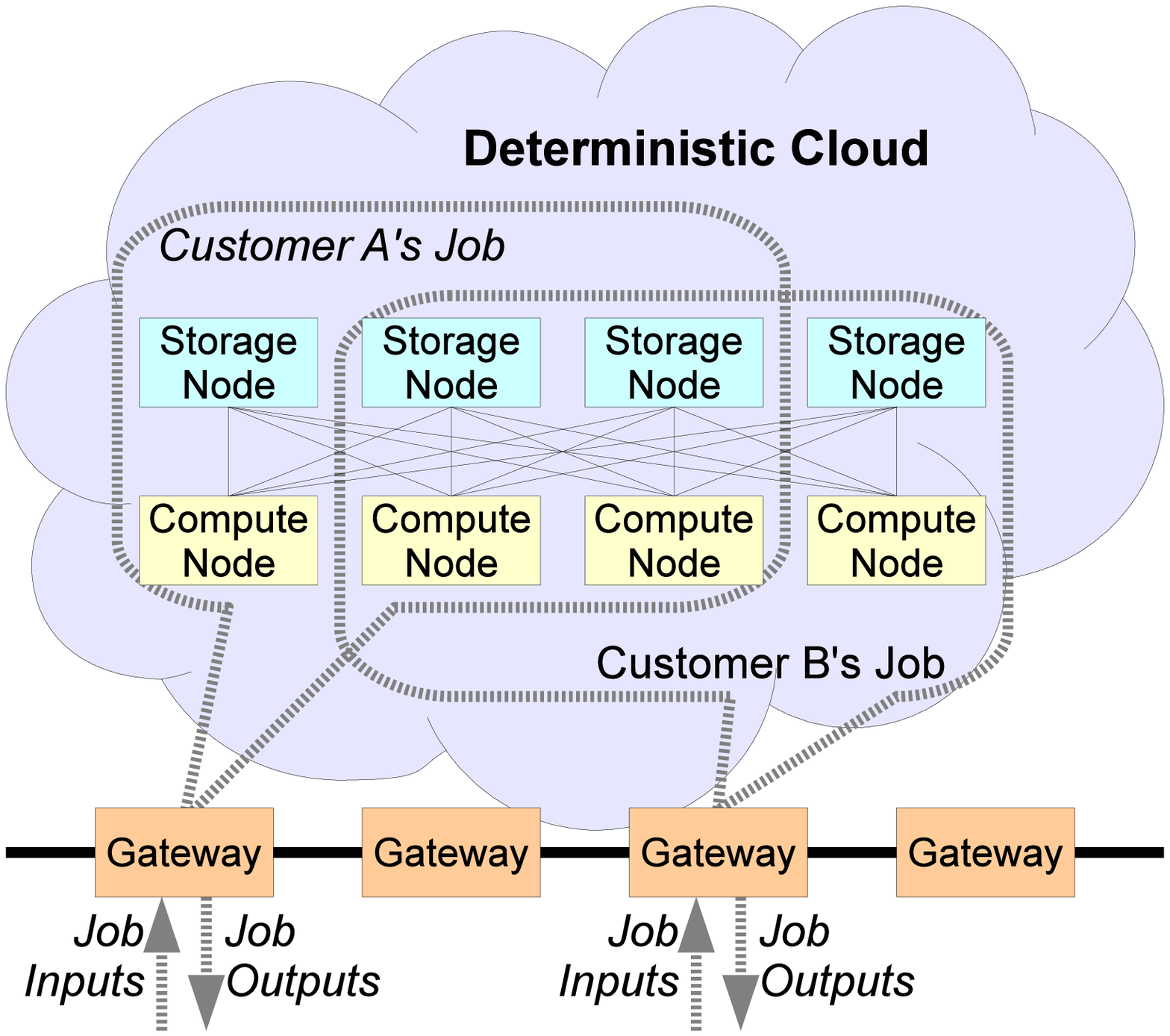}
\caption{Timing-hardened cloud architecture.
	Gateways accept requests,
	dispatch deterministic jobs into the cloud,
	then return job results that depend {\em only} on explicit job inputs,
	and not on internal timing.}
\label{fig-cloud}
\end{figure}

As illustrated in Figure~\ref{fig-cloud},
a set of gateway nodes at the cloud's boundary
accepts job requests, including any inputs the job requires.
Upon completion,
the gateway returns the job's outputs,
which depend {\em only} on explicit inputs,
and not on timings of operations within the cloud.
For each job,
the cloud provider effectively computes a {\em pure mathematical function},
whose outputs depend only on the job's explicit customer-provided inputs,
and nothing else.
The provider's cloud OS or VMM enforces this determinism,
ensuring that even malicious guest code can do nothing
to make its results depend on internal timing or other implicit inputs.

To process each job,
the provider's gateway breaks the job into smaller work units
and uses load-balancing algorithms controlled by the provider
to distribute work among cloud servers.
These servers may communicate internally while performing a job,
provided communication timing cannot affect computed results.

A customer's job may also read and write
the customer's persistent data stored in the cloud,
provided any writes remain invisible
both externally and to other jobs
until the writing job completes.
Each job in effect executes within a provider-enforced transaction.

The provider may statistically multiplex different customers' jobs freely
onto shared hardware within the cloud,
with no static partitioning or scheduling noise injection.
Provider-enforced determinism
nevertheless ensures that no timing or other nondeterministic information
leaks from one guest computation to another,
and only one unit of timing information per job leaks to the outside world:
namely the total time the job took to complete.
This remaining timing channel leaks only heavily aggregated information
that is unlikely to be easily exploitable,
and the provider can limit this timing channel's information flow rate
by returning job results to customers on a periodic schedule---%
e.g., once per millesecond, second, or minute---%
rather than immediately on job completion.
\com{
The provider may easily rate-control this timing information leak
by controlling job input and output times in the gateways,
if the timing aggregation
}

\subsection{Applicability of the Architecture}

The applicability of this cloud architecture
depends on two questions:
whether a strictly deterministic execution environment
can provide a practical programming model for cloud applications,
and whether such a deterministic environment can be efficient enough.
We address the first question here
and the second in Section~\ref{sec-impl}.

This architecture may be readily applicable
to many large, parallel, compute-bound applications
such as scientific computing, rendering, and data analysis.
Nondeterminism in parallel applications is usually undesired~\cite{
	lee06problem,bocchino09parallel},
so eliminating it benefits the developer.
The only common {\em intentional} nondeterminism in such applications
is for internal performance optimization purposes---%
e.g., distributing work items to workers
according to dynamic availability and load---%
and our architecture delegates these functions to the cloud provider.
Determinism thus simplifies the customer's programming task
by eliminating pervasive heisenbugs~\cite{lu08learning},
making all bugs reproducible~\cite{leblanc87debugging},
and offloading load-balancing responsibilities to the provider.
Applicability thus reduces to the efficiency question.

While large compute-bound applications
fit the proposed architecture most naturally,
more interactive uses may be feasible as well.
A deterministic cloud might host
interactive web applications,
for example,
as follows.
The provider's gateway nodes
act as generic front-end Web servers,
accepting HTTP requests from remote clients
and converting them into deterministic job submissions
on behalf of the web application's owner.
The gateway attaches a job creation timestamp to each job's inputs,
enabling the application to ``tell time'' at job granularity.
A job's results can request the gateway
to start a follow-up job at a future time,
enabling the web application to implement timeouts,
push notifications on persistent sockets, etc.
The remaining questions are whether
such a ``gateway-driven'' web programming model
can be made sufficiently familiar for customers implementing web applications,
and whether the provider can support job creation and dispatch
at sufficiently high rate and fine granularity
to handle customer response time requirements.
We believe both of these questions can be answered be answered positively,
the first using appropriate runtime libraries or virtualization mechanisms,
the second via efficient deterministic execution as described later.

\subsection{Life Without Timing Channels}

Our architecture requires that the provider manage
scheduling and load-balancing decisions within a cloud,
since enabling customers to do so would involve
leaking potentially sensitive timing information
into customer computations and their outputs.
An important concern is whether
the unavailability of this fine-grained internal timing information
will make it difficult for customers
to develop and optimize their parallel applications effectively:
e.g., to perform detailed profiling-based analysis of their applications,
or to implement application-specific dynamic optimizations
or load-balancing schemes within their applications.

The unavailability of fine-grained timing information to customers
may indeed present a challenge for application profiling purposes.
A customer's application need not run {\em always} or {\em only}
on a shared cloud, however.
The customer might perform development and testing
on a smaller private cloud
owned or exclusively leased by the customer.
Even after deployment,
the customer might distribute an application across
both shared and customer-private infrastructure,
giving the customer access to full timing information
on the physical machines the customer owns or has leased exclusively.

Some applications may require
dynamic, application-specific internal load-balancing algorithms
in order to perform well.
To support such applications,
a provider might allow customers to supply
application-specific scheduling or load-balancing ``plug-ins,''
as long as the provider's OS ensures that
these plug-ins can affect {\em only} the application's performance
and not its job outputs.
The provider's OS might enforce such constraints on load-balancing plug-ins
via sandboxing mechanisms
for untrusted kernel extensions~\cite{bershad95extensibility},
or by running the application's load-balancing code in user space
and using DIFC techniques~\cite{efstathopoulos05labels,zeldovich06making}
to track processes that have been ``tainted'' with timing information,
and prevent this timing information from leaking back to the customer.

\com{
We also focus on two broad classes of cloud applications:
parallel compute-bound tasks such as
scientific computing, rendering, or data analysis;
and interactive Web applications providing access to
relatively static data sets,
such as Google Maps.
Cloud applications requiring fine-grained interactions
between many concurrent users present additional challenges
that we make no attempt to address here.
}

\com{
Now that we have reviewed the difficulties
that timing attackes present to cloud computing,
this section proposes a general-purpose architecture
to address these difficulties
{\em without} undermining the economic basis for cloud computing
by giving up demand-driven statistical multiplexing of shared hardware
among multiple customers.
The intent is that a cloud provider could adopt an architecture like this one
in order to set up and sell access to a ``timing channel hardened cloud,''
which might command higher prices than a conventional cloud
due to the higher security offered.
}
\com{
Providers can offer both hardened and conventional cloud services at once,
as long as the two services run in separate physical infrastructure domains---%
on separate server clusters each with a separate physical interconnect---%
so that no high-rate timing information leaks from the hardened cloud
to the conventional one.
}

\com{
\subsection{Timing Domains}

In this architecture, the cloud provider divides up
its hardware infrastructure infrastructure into multiple {\em timing domains}
and control the flow of timing information between these domains.
As illustrated in Figure~\ref{fig-arch},
there are at least three relevant types of timing domains:
a {\em public timing domain} representing ``the Internet''
and in general all infrastructure outside of the cloud provider's control,
a {\em private timing domain} for each customer,
and a single {\em trusted, shared timing tomain}
in which ``elastic'' computing jobs run
on statistically multiplexed hardware resources.

Each timing domain could consist of a physically disjoint set of servers,
each with its own dedicated local-area interconnect,
thereby ensuring ``physically'' that timing information
produced by the internal operations of one timing domain
is not visible to software running in other timing domains.
Alternatively, the cloud provider could host multiple timing domains
on shared hardware but use demand-independent static partitioning techniques,
like those commonly used to close timing channels
in traditional high-security systems,
to isolate one timing domain from another.
Thus, the provider does {\em not} statistically multiplex
different computations from different timing domains
onto shared hardware at fine granularity,
but shifts the allocation of hardware resources from one domain to another
only at slow, administrative timescales,
regardless of whether this allocation is done via physical reconfiguration
or via static virtual resource assignments or usage schedules.

\subsection{The Trusted, Shared Timing Domain}

Since different timing domains are not statistically multiplexed,
customer-private timing domains alone do not provide
the ``elasticity'' required
to make cloud computing economically compelling:
the customer-private timing domains serve only
to provide customers with a fully functional, unrestricted computing platform
located at the provider's site.
Instead, only the trusted, shared timing domain
provides true ``elasticity'' in this architecture.
Cloud customers can submit compute jobs to be run in the trusted timing domain,
either remotely over the Internet
or locally from a customer-private timing domain.
A gateway device we call a {\em timing firewall},
which guards the border of the trusted timing domain,
receives and validates submitted compute jobs,
load-balances submitted jobs onto available servers within the trusted domain,
and once a job has completed,
returns the results to the local or remote submitter.
Thus, the timing firewall essentially acts as a front-end
to a batch processing service implemented by the hardware
in the trusted timing domain.

The trusted timing domain can run arbitrary computations,
involving use of arbitrary amounts of short- or long-term storage,
with one key restriction:
all jobs running in the trusted domain
execute under a regime of {\em system-enforced determinism}.
The virtual machine monitor (VMM)
used in the trusted domain to execute submitted jobs
does not give these guest jobs access to any ``real'' clocks or timers,
or to {\em any} nondeterministic input sources whatsoever:
the VMM ensures that the results of any compute job depend {\em only}
on the explicit inputs provided by the customer at job submission time.
That is, the trusted timing domain effectively supports only
computation of pure mathematical functions,
for which a given input always yields exactly the same output,
regardless of when or how many times the computation is run with a given input.
Any information a submitted computation may need about real-world time---%
or any other nondeterministic information such as randomness
used to seed cryptographic functions---%
must be supplied by the customer as part of the inputs at job submission time.

\subsection{Determinism Closes Timing Channels}

From a perspective of rigorous timing channel analysis,
a timing channel can occur only when two ``clocks'' are available:
a {\em reference clock}, and a clock that can be {\em modulated}
(intentionally or unintentionally)
to transmit information in a way that is observable
by comparison against the reference clock~\cite{wray91analysis}.
The classic general-purpose approach to closing timing channels,
by partitioning shared resources statically,
works by preventing one user's ``modulation''
of its portion of the shared resource from being visible to other users.
Our architecture, in contrast,
allows different users' computations
to modulate shared resources freely in the trusted domain
with no static partitioning,
but instead removes all reference clocks from the shared timing domain.

Because the trusted timing domain executes submitted jobs deterministically,
no timing information---or any new information whatsoever---%
enters a compute job between its submission and its completion times.
By the definition of determinism,
all information defining each job's output
was already present in the job's inputs (submitted code and data):
the customer could perform or repeat the same computation itself
starting from the same inputs,
and the result would be exactly the same.
If we assume
that jobs are submitted and results are produced ``atomically''---%
e.g., as if the job request and reply
each consist of only one network packet---%
then the customer obtains one and only one ``sample'' of timing information
about the trusted timing domain from the execution of a given job:
namely, the total length of time the job took to complete,
between the time the customer submitted the job
and the time it received the job's results.
This single, coarse-grained ``timing leak'' contrasts with
the myriad fine-grained timing channels that are available in current clouds
when computations running in the cloud have high-resolution reference clocks.

The cloud provider can easily tune its timing firewalls to control the rate
at which sensitive information might leak through this one timing channel.
If the firewall limits each customer to submitting one new job per second,
for example,
then a customer can obtain at most one measurement per second.
If the firewall briefly buffers results the results of completed jobs
and releases them on a fixed schedule once per second,
then the resolution of each measurement the customer obtains
is limited to one-second granularity.
Rate-limiting information leakage through this timing channel
need not limit the size or duration of jobs a customer may submit:
a given job may involve an arbitrary amount of ``work'' by the provider,
measured either in terms of real time
or usage of shared resources in the trusted timing domain.
The timing firewall can buffer large job inputs or result outputs
to make their transmission atomic for purposes of timing information flow,
and/or the provider can require the customer to break large inputs or results
into chunks for transmission across the firewall incrementally
over multiple successive jobs.

\subsection{Advantages and Tradeoffs}

The obvious advantage of this architecture
is that it provides a general-purpose approach
to address all kinds of timing information leaks
between customers in a cloud environment,
independent of the details of specific shared hardware resources
or of particular privacy-sensitive computations,
{\em without} giving up statistical multiplexing.
The architecture introduces many practical challenges, however.

One challenge is that cloud computing is highly dependent on parallelism,
and current parallel processing and thread models are highly nondeterministic.
Demonstrating that deterministic reply of multiprocessor guests
could be done at all represented an impressive feat~\cite{dunlap08execution},
but deterministic execution in this case comes at a high performance cost
that would make it infeasible for day-to-day execution in the cloud.
The next section will focus on this challenge.

A second challenge is that applications have many legitimate needs
for awareness of ``real-world time'':
network protocols need timers to interact with other machines over the network,
interactive web applications need clocks and timers
in order to interact with users, and so on.
The proposed architecture can support such applications,
provided it is feasible to separate out
the directly timing-sensitive aspects of the application's computation
into a relatively small component of the application
that can run in the customer's private timing domain
(or even elsewhere on the Internet,
such as in the JavaScript code by which a user
is interacting with the cloud-based application),
leaving the ``bulk'' of the computation capable of executing
entirely deterministically within the trusted domain.
(Of course a customer could simply run the application
mostly or completely within his private timing domain,
doing so would give up most or all of the crucial elasticity benefits
of cloud computing.)
How feasible this separation of timing dependencies will be
is certain to be highly application-specific:
it may be relatively easy for many compute-bound applications
that are naturally oriented toward batch processing,
such as scientific computing or data analysis applications;
separating highly interactive ``Web 2.0'' services this way
may prove more of a challenge, however.
We leave attempts at answering this question to future work.

\section{Provider-Enforced Determinism}

While the proposed solution to timing channels is strong and general,
its practicality hinges on two important questions:
can customers readily implement and run large parallel cloud applications
in a strictly deterministic environment,
and can these deterministic compute jobs run efficiently?
We address the first question in this section,
and the second in Sections~\ref{impl} and~\ref{sec-eval}.

\xxx{make sure other benefits of determinism get mentioned}

\section{Sources of Nondeterminism}

Cloud applications often depend on timers or other nondeterministic inputs
for many reasons,
which we divide into {\em internal} and {\em external dependencies}.
Internal dependencies
are sources of nondeterminism
that may affect the application's internal operation
but do not or should not affect the application's externally visible behavior,
whereas external dependencies are those
that the application fundamentally requires to perform its function.

\paragraph{External dependencies:}
Interactive web applications
need ``wall-clock'' timers to implement application-level network protocols,
session timeouts, real-time user interface updates, etc.
Cryptographic algorithms need sources of (ideally ``true'') randomness
with which to seed cryptographic algorithms.

Compute jobs running under provider-enforced determinism
must obtain any external nondeterministic information they depend on
as part of the explicit inputs at the beginning of a given compute job,
since they cannot obtain such inputs during the compute job.
For highly interactive or time-sensitive interactions, for example,
unshared and unrestricted servers in a customer's private timing domain
could handle the time-sensitive direct interactions with remote clients,
but delegate longer compute-bound tasks to the shared timing domain.
The private intermediary servers
could explicitly supply all nondeterministic inputs required
to each compute job,
such as the current time and random number seeds.

Alternatively, the provider's timing firewalls 
could automatically supply
some standard nondeterministic inputs,
such as the current time,
to compute jobs initiated by remote clients.
Similarly, when a job completes and delivers its results,
it could request the timing firewall 
to initiate a follow-up job automatically at some future time.
Thus, the timing firewalls could provide compute jobs
with nondeterministic inputs from ``just outside the border,''
{\em between} compute jobs though not {\em during} them.
If the provider allows job initiation every ten or hundred milliseconds,
then interactive web applications might be implementable
entirely within the shared domain.

\paragraph{Internal dependencies:}
Standard parallel and distributed programming models
are highly nondeterministic because thread and process behavior
can depend on relative execution order.
Much parallel programming research
focuses on controlling this ``incidental nondeterminism,''
which is usually unwanted~\cite{lee06problem,bocchino09parallel}.
Applications also sometimes use timing dependencies intentionally
for internal optimization:
examples are work queues and load balancers,
whose purpose is typically to assign queued tasks
to the next worker thread to come available
or the back-end server least loaded at the moment.

Provider-enforced determinism
would eliminate both desired and undesired sources of internal nondeterminism
from shared compute jobs.
On the positive side, determinism
simplifies guest application replay and debugging~\cite{
	curtis82bugnet,leblanc87debugging,king05debugging,dunlap08execution},
and facilitates accountability mechanisms that rely on deterministic replay,
such as byzantine fault tolerance~\cite{castro99practical}
and peer review~\cite{haeberlen07peerreview}.
On the negative side, determinism
eliminates the customer's ability
to implement dynamic load-balancing optimizations
within a compute job:
the customer must instead rely on
the cloud provider's virtualization infrastructure
to perform all necessary internal dynamic optimization.
\xxx{explain more?}

\subsection{Deterministic Consistency}

Although eliminating unwanted determinism from parallel computations
is clearly desirable, it is not easy or necessarily cheap.

~\cite{ford10deterministic}

}

\com{ XXX (AFA)
Question:  How exactly would this solution address the scenarios
that Ristenpart et al. sketch?  What about (say) guessing somebody's
SSH password by timing his keystrokes?  Would an SSH interaction
be relegated to a batch process in the Deterministic Cloud?  How?
}

\section{A Deterministic Cloud OS}
\label{sec-impl}

Our architecture's ``magic ingredient,'' obviously,
is provider-enforced deterministic execution.
Most cloud-oriented operating systems and virtual machine monitors
replicate the inherently nondeterministic execution model
provided by the underlying multiprocessor/multicore hardware.
Recent application-level deterministic scheduling techniques
show promise~\cite{berger09grace,bergan10coredet},
but they apply only within a process
and do not prevent a guest from intentionally escaping
its ``deterministic sandbox.''
The only system we are aware of
that enforces determinism on multiprocessor guests
does so by recording and replaying a previous (nondeterministic) execution,
and imposes a high performance cost~\cite{dunlap08execution}.

To offer evidence that the proposed architecture may be practical,
we introduce Determinator,
a novel OS that enforces determinism
on multi-process parallel computations at moderate cost,
while supporting familiar parallel programming abstractions
such as fork/join synchronization, shared memory, and file systems.
We describe Determinator from a more general perspective
elsewhere~\cite{ford10efficient},
but we briefly summarize here the aspects
relevant to timing channel control in the cloud.

Determinator is intended to supervise the compute nodes
in a cloud architecture such as that shown in Figure~\ref{fig-cloud}.
We believe cloud providers will have an incentive
to deploy deterministic compute clouds
based on an OS designed along the lines of Determinator,
because of the enhanced data privacy assurance
that a deterministic cloud could offer security-conscious customers.
Integrating Determinator into
a trusted cloud computing model~\cite{santos09towards}
could further increase both real and perceived security.

Our current priority is to demonstrate the viability
of OS-enforced deterministic execution of compute-bound jobs.
Determinator currently
provides no persistent storage,
and does not emulate hardware interfaces
or host existing operating systems,
although we intend to expand Determinator's capabilities in the future.

\com{ XXX (AFA)
Question:  Do we want the whole list of what Determinator does not do
in that second sentence?  
As a prototype, Determinator currently manages processes on a multicore PC,
but we plan to extend it to manage a cloud whose nodes host existing operating systems.
}

We now outline Determinator's basic execution environment and API,
the consistency model it uses to manage state
logically shared among parallel processes,
and how it supports both threads interacting via (logically) shared memory
and Unix-like processes interacting via a (logically) shared file system.
We make no claim that this is the ``right'' way
to implement a determinism-enforcing OS,
but merely use Determinator to explore
some key design challenges and solutions,
and how Determinator's design potentially addresses
the goal of timing-hardened cloud computing.

\subsection{Process Model}

\begin{figure}[t]
\centering
\includegraphics[width=0.47\textwidth]{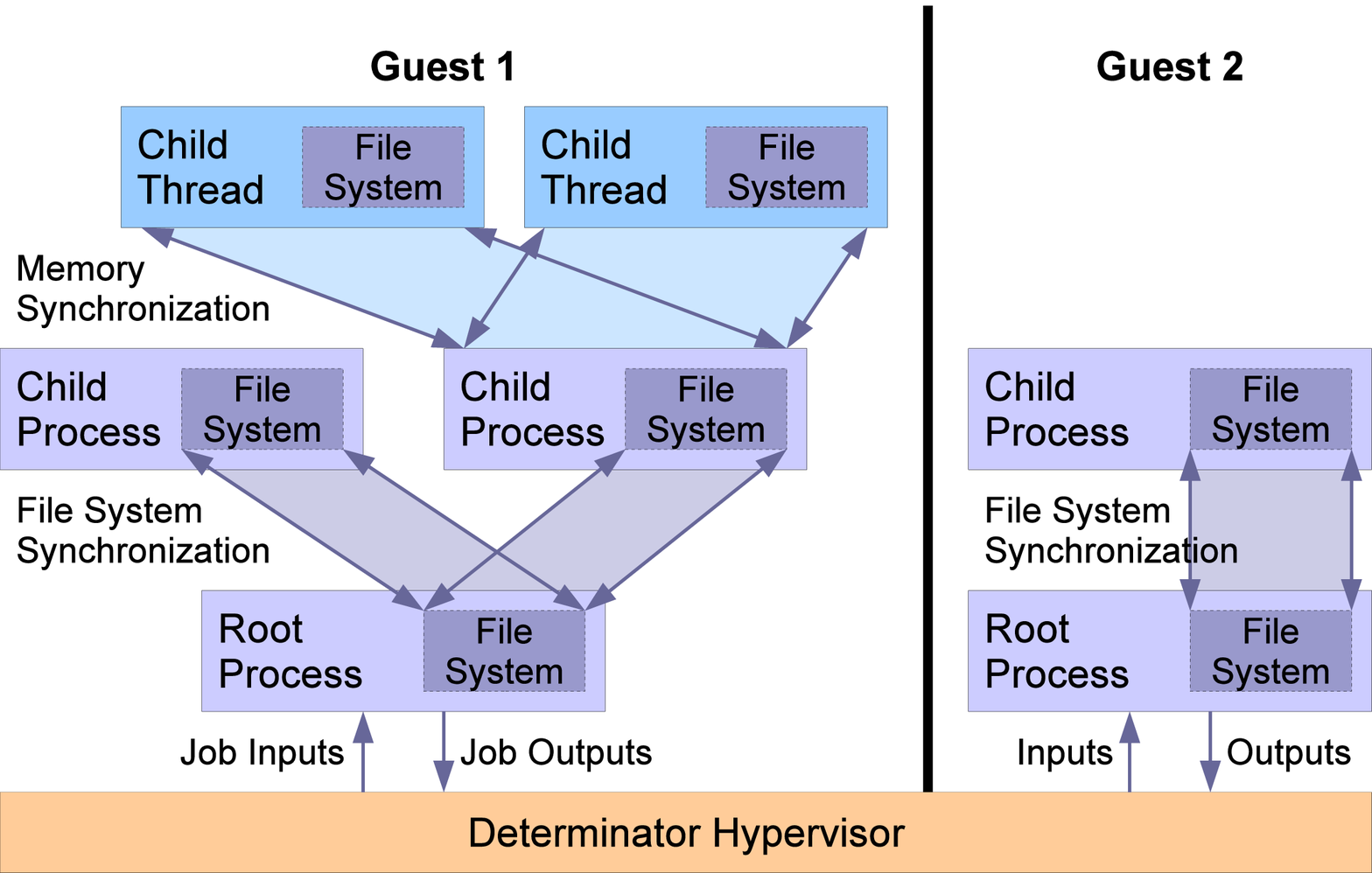}
\caption{Determinator process model.
	Each guest owns a hierarchy of processes/threads
	executing in parallel.}
\label{fig-procs}
\end{figure}

Determinator gives each guest an independent process hierarchy,
as shown in Figure~\ref{fig-procs}:
it creates a {\em root process} on behalf of the customer,
and existing processes can create new child processes.
Unlike Unix, but as in nested process models~\cite{ford96microkernels},
Determinator's hierarchy strictly constrains
process lifetime and inter-process communication.
A process cannot outlive its parent,
and a process can communicate directly
{\em only} with its immediate parent and children.

Although all guest processes can execute in parallel,
Determinator enforces determinism in two ways.
First, from the kernel's perspective,
each process is single-threaded and shares {\em no} state
with other processes.
Each process has its own registers and address space,
and processes cannot share read/write access to the same physical memory,
thereby ensuring that each process's internal execution is deterministic
as long as the processor's underlying instruction set is deterministic.
Second,
Determinator constrains the inter-process communication and synchronization
of all processes to act as a Kahn process network~\cite{kahn74semantics},
which provably yields deterministic behavior globally
in spite of parallel execution.

\subsection{Process Execution and API}
\label{sec-impl-api}

Determinator processes can have three states:
{\em runnable}, {\em stopped}, and {\em waiting}.
Runnable processes can execute concurrently with all other runnable processes,
according to a kernel-controlled scheduling policy,
but do not interact with each other while running.
(Processes could offer the kernel ``scheduling hints'' such as priorities,
which the OS might use or ignore,
but determinism precludes any explicit feedback from the OS
affecting computed results.)
A stopped process does nothing until its parent explicitly {\em starts} it.
A waiting process is blocked until a particular child stops,
at which point the waiting process becomes runnable again.

All inter-process interaction is driven by processor traps
and the kernel's three system calls: PUT, GET, and RET.
PUT waits until a designated child stops,
then copies a block of virtual memory and/or register state
into the child,
and also optionally:
(a) copies the child's entire virtual address space
	into a {\em reference snapshot} associated with the child; and/or
(b) (re-)starts the child.
GET waits until a designated child stops,
then copies or {\em merges} a block of the child's virtual memory,
and/or the child's final register state, back into the parent.
A merge is like a copy, except Determinator copies
only words that {\em differ}
between the child's current and reference snapshots
into the parent's address space,
leaving all other words in the parent untouched.
RET explicitly stops the current process,
effectively returning control to the parent.
Exceptions such as divide-by-zero in any process
have the effect of a RET,
providing the parent a status code
indicating why the child stopped.

The above interaction model ensures global determinism
because processes interact only at well-defined execution points
determined by each process's internal flow:
namely when the parent does a GET or PUT and the designated child has stopped.
The kernel gives ordinary processes no ability
to wait for ``the first child that stops,''
nor to race each other to insert or remove items
from message queues shared among multiple threads.
See the underlying formal model~\cite{kahn74semantics} for more details.

If any process contains a bug causing an endless loop,
other processes trying to synchronize with it might block forever.
To address this risk and facilitate debugging,
a processes can specify an {\em instruction limit}
when it starts a child:
the child and its descendants collectively execute
at most this many instructions before the kernel
forcibly returns control to the parent.
Counting instructions 
enables processes to regain control of errant children
without violating determinism,
and also allows processes to ``quantize'' the execution of children
and implement deterministic scheduling schemes~\cite{
	devietti09dmp,bergan10coredet}.
\com{
Some processor architectures natively support
control recovery after a precise instruction count~\cite{hp94parisc},
but this capability can be simulated on the x86~\cite{dunlap02revirt}.
}

\subsection{Emulating Logically Shared State}

Since the kernel permits processes to share no physical state,
they can communicate only by copying data via GET and PUT.
The kernel uses copy-on-write to optimize large virtual copies,
and uses similar techniques to optimize merge operations,
so merging a page that either the parent or the child have left unmodified
requires only page-level remapping.
Leveraging this efficient virtual copy primitive,
the C library linked into each process
implements {\em logical} shared state abstractions
purely in user space.
The C library emulates shared state
by treating the guest's process hierarchy like a distributed system.
Each process maintains a replica of the shared state,
and processes reconcile this state
at well-defined {\em synchronization points} during program execution,
as in replicated file systems~\cite{parker83detection}
and distributed shared memory (DSM) systems~\cite{carter91implementation}.

\paragraph{Shared File System}
Determinator's C library currently emulates the Unix file API
by reading and writing a file system image
stored in the process's own virtual memory.
(Files could alternatively be stored
in child processes not used for execution,
reducing address space usage and the danger
of wild memory writes corrupting shared files.)

The C library also implements Unix's
\verb|fork|, \verb|exec*|, and \verb|wait*| functions,
to create and execute child processes
whose virtual memory is not logically shared with the parent
but whose file system is shared.
The \verb|fork| function clones the parent process,
including file system image,
into a new child process.
The \verb|exec*| functions replace the current process,
{\em except} for its file system image,
with a new executable loaded from the file system.

The \verb|wait*| functions not only synchronize with a child process
as in Unix,
but also use file versioning~\cite{parker83detection}
to merge the parent's and child's file system changes.
The file system implements no locking or ownership,
so concurrent writes to a file cause conflicts,
which the C library detects and flags.
A conflict makes further file access attempts return errors,
until the user resolves the conflict and explicitly clears the flag
(or fixes the bug causing the conflict and reruns the job).
Concurrent writes are allowed in one case, however:
if all writes are append-only (\verb|O_APPEND|),
as with standard output or log files,
reconciliation simply collects all appends
without concern for file offsets or ordering,
yielding effects analogous to those of asynchronous appends in Unix.

\paragraph{Shared Memory}
Determinator's C library also emulates shared memory parallelism,
currently via a simple thread fork/join API.
The \verb|tfork| function clones the entire parent process, like \verb|fork|,
but \verb|tjoin| not only merges file system changes
but also merges the child's changes to regular process memory into the parent,
using the kernel's {\em merge} operation
described in Section~\ref{sec-impl-api}.
The result is a deterministic analog
of release-consistent DSM~\cite{carter91implementation}
we refer to as {\em deterministic consistency},
detailed elsewhere~\cite{ford10deterministic}.
Unlike deterministic schedulers that emulate sequential consistency
by executing threads under an artificial ``round-robin'' schedule~\cite{
	devietti09dmp,berger09grace,bergan10coredet},
deterministic consistency need not rely on speculation to achieve parallelism
and never needs to re-execute code due to misspeculation.
Deterministic consistency also makes the effects of parallel execution
not only precisely {\em repeatable}
but also more {\em predictable} to the software developer.
If two threads execute the statements $x=y$ and $y=x$ concurrently,
for example,
under deterministic consistency the result
is {\em always} to swap the values of $x$ and $y$,
whereas under deterministic schedulers the result
depends on relative code path lengths
and hence on subtle program input variations.
Determinator's runtime can also provide deterministic scheduling
for compatibility with legacy parallel code,
though this execution mode has performance and predictability costs~\cite{
	ford10efficient}.

\com{
\subsection{Resource Management}

Current prototype doesn't do demand paging at all
except for copy-on-write,
but could easily page to disk and/or over the network.

Simplistic scheduling algorithm,
but could be enhanced.
In any case, under provider's control:
guest might provide scheduling ``hints,''
but determinism precludes the kernel providing feedback
within a job
in response to those hints.

What if a job runs too long or forever?
Customer can set a time limit or instruction limit or kill it remotely.
If the limit is a deterministic quantum,
then customer can get back the partial execution state
after expiration,
for debugging etc.
}

\subsection{Implementation}

An early Determinator prototype
currently runs on the 32-bit x86 architecture,
and implements both the shared file system
and shared memory parallel APIs described above
atop the kernel's deterministic ``shared-nothing'' processes.
The prototype has no TCP/IP networking or persistent storage as yet,
and merely accepts jobs from the console.
The shared file system supports only 256 files, each up to 4MB in size,
reflecting the limitations of a 32-bit address space.
The prototype nevertheless suggests the feasibility
of providing convenient and familiar parallel programming abstractions
under a regime of kernel-enforced determinism.

\subsection{Preliminary Results}
\label{sec-eval}

To offer some evidence
that the timing-hardened cloud computing architecture proposed in this paper
may be feasible and efficient at least for some workloads,
we briefly evaluate the current Determinator prototype
using several parallel benchmarks.
We use the following benchmarks:
{\em md5} is an ``embarrassingly parallel'' brute-force MD5 password cracker;
{\em matmult} is a $1024 \times 1024$ integer matrix multiply;
{\em qsort} is a recursive parallel quicksort on an integer array;
{\em blackscholes} is a financial benchmark
from the PARSEC suite~\cite{bienia08characterization};
{\em fft} is a parallel Fast Fourier Transform
	from SPLASH-2~\cite{woo95splash2}; and
{\em lu\_cont} and {\em lu\_noncont} are LU-decomposition benchmarks
	also from SPLASH-2.
We ran all benchmarks on a 12-core (2 sockets $\times$ 6 cores),
2.2 GHz AMD Opteron PC.

\begin{figure}[t]
\centering
\includegraphics[width=0.48\textwidth]{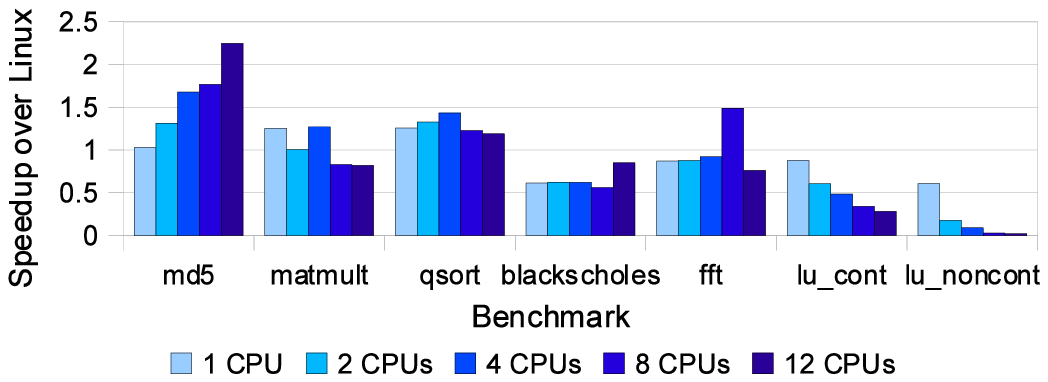}
\caption{Performance of several parallel benchmarks
	running deterministically on Determinator,
	versus nondeterministic execution on Linux.}
\label{fig-perf-bench}
\end{figure}

Figure~\ref{fig-perf-bench}
shows each benchmark's performance running deterministically on Determinator,
normalized to nondeterministic execution performance on Ubuntu Linux 9.10,
using 1--12 CPU cores.
Coarse-grained parallel benchmarks
such as {\em md5}, {\em matmult}, and {\em qsort},
which perform a substantial amount of computation
between inter-thread synchronization events,
consistently run nearly as fast and sometimes faster on Determinator
compared with Linux.
The {\em md5} benchmark surprisingly scales
much better on Determinator than on Linux,
achieving more than $2\times$ speedup over Linux on 12 cores;
we have not yet determined the precise cause of this performance increase
but suspect bottlenecks in Linux's thread system~\cite{behren03}.
The {\em blackscholes} benchmark is also ``embarrassingly parallel,''
but our port of this benchmark uses deterministic scheduling
for compatibility with the pthreads API,
incurring a constant performance overhead~\cite{ford10efficient}.
The more fine-grained SPLASH-2 benchmarks exhibit higher performance costs
on Determinator due to their more frequent inter-thread synchronization.

\com{
\begin{figure}[t]
\centering
\includegraphics[width=0.48\textwidth]{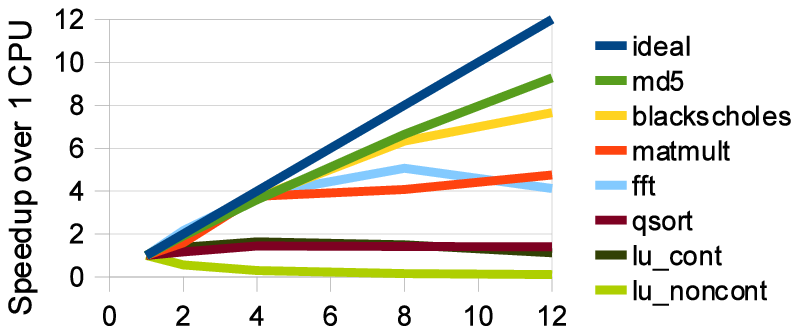}
\caption{Matrix multiply speedup with 2 and 4 CPUs,
	Linux pthreads versus Determinator threads,
	with a variety of matrix sizes.}
\label{fig-perf-speedup}
\end{figure}
}

We also examined whether we could more easily reduce (though not eliminate)
timing information leaks in stock Linux kernels,
simply by removing access to accurate timers
in both the kernel and applications.
Disabling these high-resolution timers does not prevent processes
from creating {\em ad hoc} timers via parallel threads,
of course,
as discussed in Section~\ref{sec-motiv}
and illustrated in Figure~\ref{fig-refclock}.
Nevertheless,
to test the effect of timer unavailability on a stock OS,
we compiled
the Linux kernel and applications to eliminate use of
cycle counting instructions such as \texttt{rdtsc}
and high-resolution timers.
Interestingly, we found that the throughput of the Apache web server
under load dropped by about 20\% compared to the unmodified case,
because web server and the kernel TCP/IP stack
rely on high-resolution timers for estimating client
latency, cache sizes, etc.
This result suggests that there are no simple
workarounds to close timing channels
while delivering high throughput.

TCP's dependency on high-resolution timers
does not present an immediate problem in our proposed cloud architecture,
as long as TCP is implemented in a provider-controlled kernel or VMM:
the provider's kernel is trusted and can use high-resolution timers.
Dependencies on high-resolution timers in application-level suites
such as Web services, however,
are likely to present a pragmatic challenge
when run under any timing channel control mechanism;
we leave further evaluation of these challenges to future work.

\section{Related Work}
\label{sec-related}

Timing channels are well-studied~\cite{
	kemmerer83shared,wray91analysis},
but only recently examined in the cloud context~\cite{
	ristenpart09cloud,chen10side}.
Most proposed solutions to recent cache-based attacks~\cite{
	aciicmez07predicting,aciicmez07yet,percival05cache,wang06covert}
involve cache partitioning~\cite{kong08deconstructing},
requiring hardware modifications and decreasing performance.
Specific algorithms may be hardened~\cite{vuillaume06flexible},
but the only known general solution---%
resource partitioning---%
limits statistical multiplexing and undermines the cloud business model.

Deterministic execution has been used for other purposes
such as replay debugging~\cite{leblanc87debugging}
and intrusion analysis~\cite{dunlap02revirt},
and its benefits for parallel programming
are well-recognized~\cite{lee06problem,bocchino09parallel}.
Parallel languages such as SHIM~\cite{edwards08programming}
and DPJ~\cite{bocchino09parallel}
provide deterministic programming models for these reasons,
but they cannot run legacy or multi-process parallel code.
User-level deterministic schedulers~\cite{
	berger09grace,bergan10coredet}
can provide determinism within one well-behaved process,
but cannot supervise multiple interacting processes
or prevent misbehaved applications
from escaping the deterministic environment.

Cloud providers must be able to {\em enforce} determinism in guests
in order to eliminate timing channels using our architecture.
The only system we know of that can enforce determinism
on multiprocessor guests is SMP-ReVirt~\cite{dunlap08execution}.
While impressive,
SMP-ReVirt is designed to replay prior nondeterministic executions,
rather than to execute guests deterministically ``from the start,''
and its performance cost is too high for everyday use.


\section{Conclusion}
\label{sec-concl}

We have proposed a new, general approach
to combating timing channels in clouds
via provider-enforced deterministic execution.
The key benefit of this approach is that it eliminates
the exploitability of {\em all} timing channels internal to a cloud,
independent of the type of resource manifesting the channel,
without undermining the cloud's elasticity
through resource partitioning.
Preliminary results from our determinism-enforcing OS
suggest that such a timing-hardened architecture
may be feasible and efficient at least for some applications,
but many questions remain.
Can such an architecture support
fine-grained parallel applications,
interactive Web applications,
transactional storage- or communication-intensive applications?
Can it offer cloud customers
a rich and convenient, yet efficient,
programming model in which to express such applications deterministically?
Can deterministic clouds reuse legacy software and operating systems?
Only further exploration will tell.

\com{
\subsection*{Acknowledgments}

Zhong

funding?
}

\begin{footnotesize}
\bibliography{os}
\bibliographystyle{abbr}
\end{footnotesize}

\end{document}